\documentclass[aps,prb,twocolumn,showpacs]{revtex4}
\usepackage{graphicx}
\usepackage{epsfig}
\usepackage{amsmath}
\usepackage{amssymb}
\usepackage{amsfonts}%
\begin{document} 
\title{Successes and failures of Bethe Ansatz Density Functional Theory}
\author{Stefan Schenk}
\author{Michael Dzierzawa}
\author{Peter Schwab}
\author{Ulrich Eckern}
\affiliation{Institut f\"ur Physik, Universit\"at Augsburg, 86135 Augsburg, Germany}

\begin{abstract} 
The response of a one-dimensional fermion system is investigated
using Density Functional Theory (DFT) within the Local Density Approximation
(LDA), and compared with exact results. It is shown that DFT-LDA
reproduces surprisingly well some of the characteristic
features of the Luttinger liquid, namely the vanishing spectral weight
of low energy particle-hole excitations, as well as the dispersion of
the collective charge excitations.
On the other hand, the approximation fails, even qualitatively, for
quantities for which backscattering is important, i.e., those quantities
which are crucial for an accurate description of transport. In particular,
the Drude weight in the presence of a single impurity is discussed.
\end{abstract} 

\date{\today} 
\maketitle 

\section{Introduction}

Density Functional Theory (DFT) is the most efficient and powerful tool for
determining the electronic structure of solids.
While originally developed for continuum electron systems with Coulomb
interaction,\cite{hohenberg1964,kohn1965}
DFT has also been applied to lattice models, such as the Hubbard
model.\cite{gunnarsson1986,schonhammer1987,schonhammer1995,lima2003}
One goal of these studies was to develop 
new approaches to correlated electron systems:
lattice models often allow for exact solutions
-- either analytically or based on numerics -- which hence can serve as
benchmarks for assessing the quality of approximations that usually have to
be made when using DFT. 
Very popular in solid state applications is the Local Density Approximation (LDA) 
where the exchange-correlation energy of the inhomogeneous system
under consideration is constructed via a local approximation from the
homogeneous electron system. 
Amongst others, LDA has been applied to study ultracold fermions in one-dimensional
optical lattices,\cite{xianlong2006} Friedel oscillations in one-dimensional
metals,\cite{alcaraz2007} the Mott gap in the Hubbard model,\cite{lima2002}
and quantum spin chains.\cite{alcaraz2007}

For small systems and not too strong interactions, LDA in most
cases produces reasonable results -- which can be obtained with much
less numerical effort than needed when using more accurate methods like exact
diagonalization or the density matrix renormalization group. This led to the
hope that the LDA may serve as a useful tool for large systems,
where the numerical effort for the more accurate methods is too 
expensive.

After recalling in the next section (Sect.~II) the theoretical background for
the Bethe ansatz LDA, we will study in detail the LDA solution of
spinless fermions in one dimension. In Sect.~III we determine the
charge susceptibility and discuss, in particular, questions of stability as 
well as the static and the dynamic response. Then (Sect.~IV) we study the
Drude weight (which can be related to the conductivity) in the presence of
a single impurity, and present our conclusions in the final Sect.~V. 
From the size-dependence of our results, we find that for large systems, LDA
predictions are qualitatively incorrect even for weak interaction.

\section{Formalism}
\label{secDFT}
We consider a one-dimensional model of spinless fermions described by the Hamiltonian
\begin{equation} \label{eq1}
\hat H = - t \sum_i (\hat c_i^+ \hat c_{i+1} + {\rm h.c.}) + 
V \sum_i \hat n_i \hat n_{i+1} + \sum_i v_i \hat n_i
\end{equation}
where $\hat c_i^+ (\hat c_i)$ creates (annihilates) a fermion at site $i$, 
$t$ is the hopping parameter,
$V$ the nearest-neighbor interaction, and $v_i$ an arbitrary local potential.
The lattice consists of $L$ sites (the lattice constant $a$ is set to one), 
and periodic boundary conditions are assumed.
%The homogeneous model, without external potential, has been solved 
%using the Bethe ansatz.\cite{yang1966}

The lattice version\cite{gunnarsson1986} of
DFT relies on the fact that there is a one-to-one correspondence
between the potentials $\{v_i\}$ and the groundstate expectation values of the site
occupations $\{n_i\}$.
Therefore it is -- in principle -- possible to express all quantities that can be
obtained from the groundstate wave function as a function (or
functional in the continuous case) of the densities.
The site occupations as a function of the potentials can, of course, be found
from derivatives of the groundstate energy with respect to the
local potential,
\begin{equation} 
n_i = \frac{\partial E_0}{\partial v_i} \; .
\end{equation}
On the other hand,
in order to determine the potentials from the densities
it is convenient to define the function
\begin{equation} \label{eq2}
F(\{n_i\}) = \min_{\Psi\rightarrow\{n_i\}} \langle\Psi|\hat T + \hat V|\Psi\rangle
\end{equation}
where $\Psi\rightarrow\{n_i\}$ indicates that the minimization is constrained to
such wave functions $\Psi$ that yield the given site occupation, i.e.,
$\langle\Psi| \hat n_i |\Psi\rangle = n_i$. Here
$\hat T$ and $\hat V$ are the kinetic and interaction part of 
the Hamiltonian (\ref{eq1}),
respectively.
The groundstate energy is obtained by
minimizing the function $E(\{n_i\}) = F(\{n_i\}) + \sum_i v_i n_i$ with
respect to $n_i$. This yields the condition
\begin{equation} \label{eq3}
\frac{\partial F}{\partial n_i} + v_i = 0
\end{equation}
which, of course, is purely formal unless $F$ or at least a reasonable 
approximation for it is available.

A major step towards the practical implementation of DFT was the idea of Kohn and
Sham\cite{kohn1965} to employ a non-interacting auxiliary Hamiltonian $\hat H^s$
in order to calculate the groundstate density profile.
In the present case
\begin{equation} \label{eq4}
\hat H^s = \hat T + \sum_i v_i^s \hat n_i
\end{equation}
where the potentials $v_i^s$ have to be chosen such that in the groundstate of $\hat H^s$
the site occupations $n_i$ are the same as in the interacting model.
Performing the same steps as before, one obtains the conditions
\begin{equation} \label{eq5}
\frac{\partial F^s}{\partial n_i} + v^s_i = 0 \; .
\end{equation}
Combining Eqs.\ (\ref{eq3}) and (\ref{eq5}) yields
\begin{equation} \label{eq6}
v_i^s = v_i + \frac{\partial}{\partial n_i}(F-F^s) =: v_i + v_i^{\rm H} + v_i^{\rm xc}
\end{equation}
where $v_i^{\rm H} = V(n_{i+1}+n_{i-1})$ is the Hartree potential, and
$v_i^{\rm xc}$ is the so-called exchange-correlation potential.
The minimization problem of DFT is thus mapped onto the diagonalization of
the one-particle Hamiltonian $\hat H^s$ supplemented with the self-consistency
condition (\ref{eq6}). However, there remains
the problem of finding a practical approximation
for the exchange-correlation potential $v_i^{\rm xc}$.
Most DFT studies of lattice models have so far relied on the LDA
where the groundstate energy density $\epsilon_i$
of the inhomogeneous system is approximated
by the energy density of a homogeneous system at the same density.
In the present case this quantity can be
calculated from the Bethe ansatz equations;\cite{yang1966} hence 
\begin{equation} \label{eq7}
[v_i^{\rm xc}]_{\rm LDA} = 
\frac{\partial }{\partial n_i}\left(\epsilon^{\rm BA}(n_i)-\epsilon^{\rm H}(n_i)\right)
\end{equation}
where $\epsilon^{\rm BA}(n)$ is the Bethe ansatz energy per site of a 
homogeneous system with particle density $n$, 
and $\epsilon^{\rm H}(n)$ the corresponding energy density in Hartree
approximation.

\section{Susceptibility}
In order to assess
the results based on LDA and to discuss their validity, it is appropriate
to recall first the phase diagram of the model under consideration.
In the homogeneous case and away from half filling one finds for all
values of the interaction parameter $V/t > -2  $ a Luttinger liquid phase,
i.e., there is no long range charge order and the low energy
excitations are gapless collective charge excitations.
At half filling and for $ V/t>2 $ the model exhibits 
long range charge order, and a charge gap opens.
Figure 1 shows the exchange-correlation potential $v^{\rm xc}(n)$ 
obtained from Bethe ansatz, compare (\ref{eq7}), for
several values of the interaction strength $V$.
Due to particle hole symmetry, we have $v^{\rm xc}(1-n) = - v^{\rm xc}(n)$.
Furthermore, for $V/t > 2$ there is
a discontinuity at $n = 1/2$, related to the opening of the charge gap. 

%%%%%%%%%%%%%%%%%%%%%%%%%%%%%%%%%%%%%%%%%%%%%%%%%%%%%%%%%%%%%%%
\begin{figure}
\includegraphics[width=0.44\textwidth]{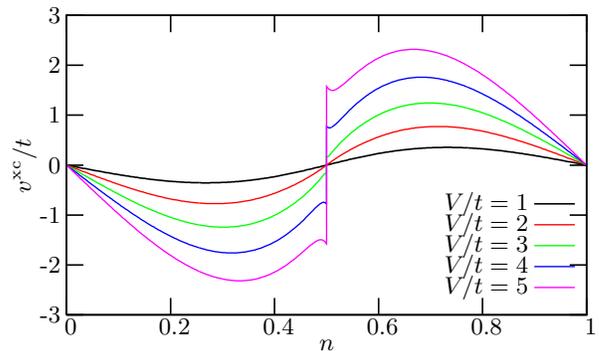}
\caption{\label{fig1}(Color online) Exchange correlation potential $v^{\rm xc}$ 
of the one-dimensional spinless fermion model as function of the density $n$
for several values of the nearest-neighbor interaction $V$.}
\end{figure}
%%%%%%%%%%%%%%%%%%%%%%%%%%%%%%%%%%%%%%%%%%%%%%%%%%%%%%%%%%%%%%%

%%%%%%%%%%%%%%%%%%%%%%%%%%%%%%%%%%%%%%%%%%%%%%%%%%%%%%%%%%%%%%%
\begin{figure}
\includegraphics[width=0.44\textwidth]{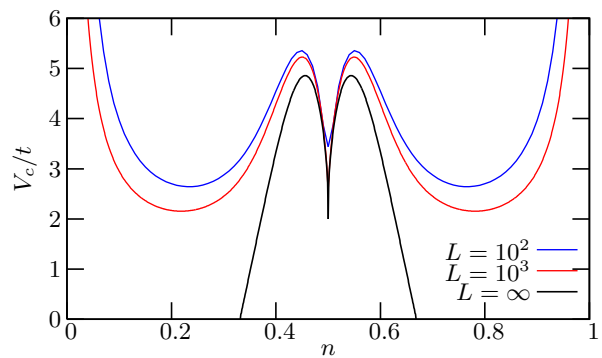}
\caption{\label{fig2}(Color online) Critical interaction strength $V_c$ above which the LDA
susceptibility is negative, indicating an instability of the system
with respect to charge ordering. 
In the infinite system, the stable region is localized near half
filling, from $n_c \approx 0.331$ to $1-n_c$.
For finite system size and weak interaction, LDA is stable for all densities.}
\end{figure}
%%%%%%%%%%%%%%%%%%%%%%%%%%%%%%%%%%%%%%%%%%%%%%%%%%%%%%%%%%%%%%%

\subsection{Stability}
First we study the stability of the homogeneous LDA solution
by considering the charge susceptibility $\chi(q)$.
Generally, the susceptibility can be expressed as
\begin{equation} \label{eq8}
\chi(q) = \frac{\chi_0(q)}{1 + [V(q) + f^{\rm xc}(q)]\chi_0(q)}
\end{equation}
where $q$ is the wavevector,
$f^{\rm xc}(q)$ the Fourier transform of $f_{i-j}^{\rm xc} = \partial v_i^{\rm xc}/\partial n_j$,
and $V(q) = 2V\cos q$;
$\chi_0(q)$ is the static susceptibility of the auxiliary system, given by ($L\to\infty)$
\begin{equation} \label{eq9}
\chi_0(q) = \frac{1}{4\pi t \sin(q/2)}\ln 
\left|\frac{\sin(q/2) + \sin k_F}{\sin(q/2) - \sin k_F}\right|
\end{equation}
%
%Within LDA the function $f^{\rm xc} \to f^{\rm xc}_{\rm LDA}$ is independent of
%$q$.
where $k_F$ is the Fermi wavevector.
The stability boundary of the homogeneous density profile is determined by the
condition that the static susceptibility becomes infinite and changes sign; 
this happens whenever the denominator in (\ref{eq8}) vanishes, i.e., for
$V(q) + f^{\rm xc}(q) = - \chi_0^{-1}(q)$. 
Due to the logarithmic divergence of $\chi_0(q)$ for $q \rightarrow 2 k_F$
this is equivalent to the condition that $V(2k_F) + f^{\rm xc}(2k_F)$ changes
sign. Notice that within LDA the function $f^{\rm xc}(q) \to
f^{\rm xc}_{\rm LDA}$ is independent of $q$.
Figure 2 shows the region of stability in the $n$-$V$-plane obtained within LDA
both for the infinite system, and for finite systems of length $L = 100$ and $L=1000$,
respectively.
For $L \to \infty$ only systems with density near $1/2$ and not too
strong interaction are stable, further away from half filling the
homogeneous solution is unstable for arbitrarily weak interaction.

For an accurate determination of the phase boundary we investigate
the weak interaction case in more detail.
We find
\begin{equation} \label{eq10}
f^{\rm xc}_{\rm LDA} = \frac{\partial^2}{\partial n^2}(\epsilon^{\rm BA} - \epsilon^{\rm H}) 
= - V(2 k_F) + {\cal O}(V^2)
\end{equation}
so that in first order in the interaction no conclusion about the
stability can be drawn: 
the second order correction to the groundstate energy, $\epsilon_2$, is needed.
Numerically we find that its second derivative with respect to the density,
$\epsilon_2^{\prime\prime}(n)$,
changes sign at $n_c \approx 0.331$, thus limiting the range of stability to $n_c < n < 1-n_c$
at weak coupling.
This result should be contrasted with the Hartree approximation ($f^{\rm xc}=0$) where the 
homogeneous system is stable only for $V(2k_F) > 0$, i.e., below quarter and above
three quarter filling, and with the exact groundstate where a charge instability 
of the homogeneous system occurs only at half filling for $V > 2t$.

There are, however, very pronounced finite-size effects that strongly enlarge
the actual region of stability within LDA.
Since $\chi_0(2 k_F)$ diverges only logarithmically with system size $L$,
the critical interaction strength approaches zero very slowly,
$V_c(L) \sim 1/\sqrt{\ln L}$. As a consequence, for finite systems and from
weak to intermediate interaction strength the homogeneous LDA 
solution is stable for all densities, as can be seen
in Fig.\ 2 for $L = 100$ and $L = 1000$.

\subsection{Static response}
Here we investigate the static susceptibility, i.e., its $q$-dependence, in
more detail.  In Fig.\ 3 we show $\chi^{\rm LDA}(q)$ for $V/t = 1$ 
in comparison with the exact susceptibility
obtained from numerical diagonalization of small systems.
As to be expected, in the long wavelength limit, $q \rightarrow 0$, 
perfect agreement is found. Technically, there is a
cancellation between the susceptibility $\chi_0^{-1}(0) = 2 \pi t \sin k_F$
and the second derivative of the Hartree energy $\epsilon^H = -(2t/\pi) \sin k_F + V n^2$
with respect to $n = k_F/\pi$. Therefore
\begin{equation} \label{eq11}
\chi^{\rm LDA}(q\rightarrow 0) = \left(\frac{\partial^2 \epsilon^{\rm BA}}{\partial n^2}\right)^{-1} = \frac{1}{L}\frac{\partial N}{\partial \mu} 
\end{equation}
which is the exact uniform susceptibility of the interacting system.
Unfortunately, already the next to leading contribution, $\sim q^2$,
is not obtained correctly within LDA.
At half filling the discrepancy between the LDA
susceptibility and the exact one becomes more and more pronounced for
$q \to 2k_F = \pi$. At  $q = 2k_F$ the exact susceptibility
increases with the system size (not shown in the figure)
and diverges with a power law, while in LDA there is only a cusp.
The cusp value itself remains finite and approaches
$\chi^{\rm LDA}(\pi) = 1.668/t$ for $L\rightarrow \infty$.

At quarter filling $\chi^{\rm LDA}(q)$ is very close to the exact
susceptibility for $q < 2k_F$, while for $q > 2k_F$ there is a clear discrepancy.
For $q = 2k_F = \pi/2$ the exact result again is strongly size-dependent and diverges 
for $L \rightarrow \infty$,  while within LDA the susceptibility diverges
already at a finite systems size, since  at quarter filling
one is already outside the range of stability of LDA.

%%%%%%%%%%%%%%%%%%%%%%%%%%%%%%%%%%%%%%%%%%%%%%%%%%%%%%%%%%%%%%%
\begin{figure}
\includegraphics[width=0.44\textwidth]{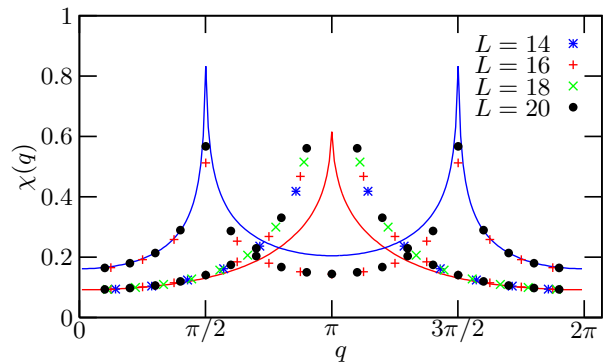}
\caption{\label{fig3}(Color online) Static susceptibility $\chi(q)$ 
(in units of $t^{-1}$) in LDA for $V/t = 1$ at half filling
($L=202$, one-peak curve) and quarter filling ($L = 204$, two-peak curve). 
The symbols are results from exact
diagonalization for systems of up to $L=20$ sites.}
\end{figure}
%%%%%%%%%%%%%%%%%%%%%%%%%%%%%%%%%%%%%%%%%%%%%%%%%%%%%%%%%%%%%%%

%%%%%%%%%%%%%%%%%%%%%%%%%%%%%%%%%%%%%%%%%%%%%%%%%%%%%%%%%%%%%%%
\begin{figure}
\includegraphics[width=0.44\textwidth]{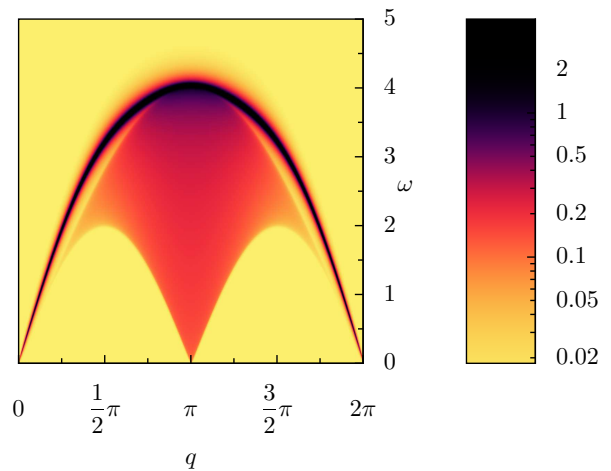}
\caption{\label{fig4}(Color online) Imaginary part of the dynamical 
susceptibility $\chi^{\rm ALDA}(q,\omega)$ (in units of $t^{-1}$).}
\end{figure}
%%%%%%%%%%%%%%%%%%%%%%%%%%%%%%%%%%%%%%%%%%%%%%%%%%%%%%%%%%%%%%%

\subsection{Dynamic response}
DFT as presented in Sect.~\ref{secDFT} is a groundstate theory.
However, a time-dependent generalization\cite{runge1984} of DFT is
available which allows to calculate frequency-dependent
response functions.\cite{gross1985}
The dynamic susceptibility of the homogeneous system is of the same form 
as Eq.\ (\ref{eq8}) with the only
differences that $\chi_0(q)$ has to be replaced by $\chi_0(q,\omega)$, and
$f^{\rm xc}(q)$ by $f^{\rm xc}(q, \omega)$. In a simple approximation, called
adiabatic local density approximation\cite{zangwill1980} (ALDA), one neglects this
frequency dependence and uses the function $f^{\rm xc}$ already known from
LDA, $f^{\rm xc}(q,\omega) \to f^{\rm xc}_{\rm LDA} $. The corresponding 
approximation for the
susceptibility is denoted $\chi^{\rm ALDA}(q,\omega)$.

%%%%%%%%%%%%%%%%%%%%%%%%%%%%%%%%%%%%%%%%%%%%%%%%%%%%%%%%%%%%%%%
\begin{figure}
\includegraphics[width=0.44\textwidth]{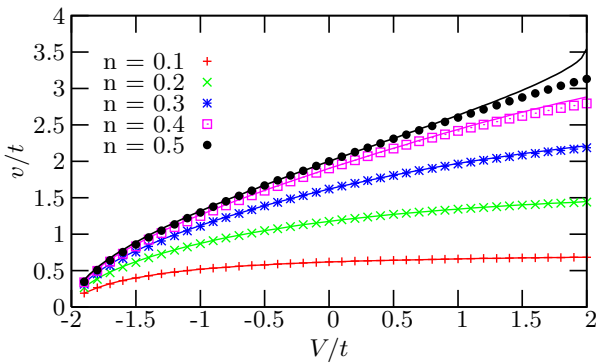}
\caption{\label{fig5}(Color online) Charge velocity $v$ as function of the interaction strength $V$ for
densities $n = 0.1$, 0.2, 0.3, 0.4, 0.5 (from bottom to top).
Exact values from Bethe ansatz (solid lines) in comparison with the 
results obtained within
ALDA (symbols) from Eq.\ (\ref{eq14}).}
\end{figure}
%%%%%%%%%%%%%%%%%%%%%%%%%%%%%%%%%%%%%%%%%%%%%%%%%%%%%%%%%%%%%%%

%
Figure 4 shows the imaginary part of 
$\chi^{\rm ALDA}(q,\omega)$ in the $q$-$\omega$-plane for
a half-filled system and $V/t = 1$. A continuum of excitations in the
frequency range $2t|\sin q| < \omega < 4t \sin(q/2)$ is apparent, which can be
identified with the particle-hole continuum. 
Note that the spectral weight of the particle-hole excitations vanishes for $q \to 0$,
a feature -- expected for a Luttinger liquid -- which is reproduced 
surprisingly well in ALDA.

Above this continuum we find a well-defined
branch of collective excitations with linear dispersion for small $q$.
Analytically, the dispersion of the collective excitations can be obtained from the singularities
of the susceptibility, i.e., by considering the zeros of the denominator of $\chi(q,\omega)$.
In the low frequency and small wavevector limit the susceptibility 
agrees with the Luttinger liquid result
\begin{equation} \label{eq13}
\chi (q,\omega) \approx \frac{1}{L} \frac{\partial N }{\partial \mu }
\frac{(q v)^2}{ (q v)^2 - \omega^2}
\end{equation}
where $v$ is the velocity of the collective excitations.
Within the adiabatic local density approximation the velocity is given by
\begin{equation} \label{eq14}
v_{\rm ALDA} = v_F \sqrt{1 + \frac{2V + f^{\rm xc}_{\rm LDA}}{\pi v_F}} 
\end{equation}
where $v_F = 2t \sin k_F$ is the Fermi velocity.
To linear order in the interaction, $v_{\rm ALDA}$ agrees with the
exact result.
%
%\begin{equation} \label{eq15}
%v_{\rm ALDA} = v_F \left(1 + \frac{V}{\pi v_F}(1 - \cos 2k_F)\right)
%\end{equation}
% 
In Fig.\ 5 we compare $v_{\rm ALDA}$ with the exact value, obtained from
Bethe ansatz, for various interaction strengths and densities. 
For small densities, there is nearly perfect agreement
between the two values.  
The largest deviation occurs at half filling at the critical point $V = 2t$ where the 
error is $2/\sqrt{\pi} - 1$, which is about 13~\%.
We want to emphasize that within the random phase approximation, i.e., neglecting the 
ALDA correction factor $f^{\rm xc}$ in
Eq.\ (\ref{eq14}), one never obtains the correct charge velocity, except for $V = 0$.
From the discrepancy between $v_{\rm ALDA}$ and the exact value one
concludes that the frequency and wavevector dependent function
$f^{\rm xc}(q, \omega)$ is non-analytic in the $q , \omega \to 0$ limit.

%%%%%%%%%%%%%%%%%%%%%%%%%%%%%%%%%%%%%%%%%%%%%%%%%%%%%%%%%%%%%%%
\begin{figure}
\includegraphics[width=0.44\textwidth]{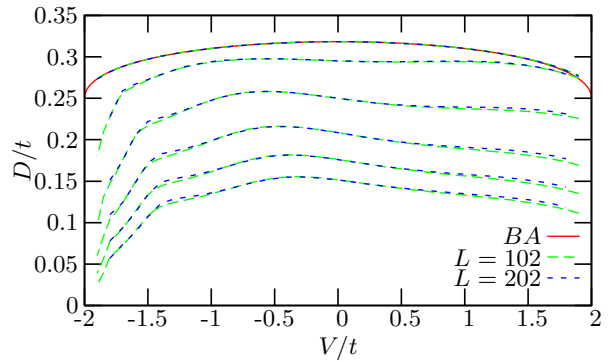}
\caption{\label{fig6}(Color online) Drude weight $D$ for a half-filled
system with a single impurity as function of
the interaction strength $V$, for several values of the impurity strength
$v_{\rm imp}/t = 0$, 1, 2, 3, 4, 5 (from top to bottom). The solid curve is 
the Bethe ansatz
result for the clean system, the long- and short-dashed curves are 
LDA results for $L = 102$ and $L = 202$, respectively.}
\end{figure}
%%%%%%%%%%%%%%%%%%%%%%%%%%%%%%%%%%%%%%%%%%%%%%%%%%%%%%%%%%%%%%%

\section{Single impurity}
Finally, we consider the spinless fermion model for a non-zero potential ${v}_i$. 
As an important example in this context, we consider the case of a 
single impurity, i.e., ${v}_i
= v_{\rm imp}$ at the impurity site and ${v}_i =0 $ elsewhere. 
It is well known that in one-dimensional interacting systems
the reflection and transmission probabilities for scattering at
an impurity are strongly 
renormalized.\cite{kane1992,matveev1993,schmitteckert1998} 
Here we address the question
whether this renormalization is captured by the LDA -- with negative conclusion.

Let us first consider the simple picture for the origin of the
renormalization that has been developed by Matveev et al.\cite{matveev1993}
Around an impurity, the density is disturbed, and in the presence of
electron-electron interaction this modulation (Friedel oscillations) leads to
additional scattering. In particular, the Friedel
oscillations are the origin of enhanced backscattering. 
To linear order in the
interaction, the correction to the transmission probability for a 
wavevector $q$ close to $k_F$ is given by\cite{matveev1993}
\begin{equation} \label{eq15}
\delta {\cal T} = - 2 \alpha {\cal T}_0(1- {\cal T}_0 )  
\ln \left( \frac{1}{|q-k_F|} \right) \; ,
\end{equation}
where ${\cal T}_0$ is the bare value; the
dimensionless parameter $\alpha$ characterizes the interaction strength.
It is given by the sum of a Hartree and an exchange contribution,
$\alpha = \alpha_{\rm H} + \alpha_{\rm x}$, with 
$\alpha_{\rm H} =- V(2k_F)/2 \pi v_F$ and $\alpha_{\rm x} = V(0)/2 \pi v_F$.
By summation of the leading divergencies to all orders in the interaction
using a renormalization group approach, it is found that even for a
weak defect the transmission approaches zero as $q \to k_F$ (repulsive
interaction).
Repeating the arguments leading to Eq.~(\ref{eq15}) 
within DFT and for a weak impurity, we
find $\alpha  \to - [V(2 k_F) + f^{\rm xc}(2k_F)]/ 2 \pi v_F$. Since
$f^{\rm xc}_{\rm LDA}= - V(2k_F)$, this singular correction
to the transmission is zero, i.e., ${\cal T}$ is {\em not}
renormalized in DFT-LDA.

To substantiate this finding numerically we calculate the Drude
weight for the single-impurity case. 
The Drude weight, $D$, is defined as the response of the system 
to a change of boundary conditions according to
\begin{equation} \label{eq16}
D = \frac{L}{2} \left.\frac{{\rm d}^2 E}{{\rm d}\varphi^2}\right|_{\varphi=0} \; ,
\end{equation}
where $E(\varphi)$ is the groundstate energy.
The parameter $\varphi$ characterizes the twist in the boundary conditions:
$\varphi = 0$ corresponds to periodic, and
$\varphi = \pi$ to antiperiodic boundary conditions.\cite{schmitteckert1998,eckern1995}
The Drude weight is closely related to the transmission through
the defect, and in the non-interacting system -- where ${\cal T}$ is not
renormalized -- the size dependence of $D$ is negligible.
In the interacting system, on the other hand, the transmission coefficient for   
$(q-k_F)\approx v_F/L$  is relevant as discussed above. Correspondingly, the 
Drude weight increases
(decreases) algebraically with system size for repulsive (attractive) 
interaction.\cite{kane1992,matveev1993,schmitteckert1998}

In Fig. 6 we present our LDA results for the Drude weight at half filling, 
for different system sizes ($L = 102$ and $L = 202$)
and different values of the impurity strength. Unlike the (numerically) exact
results,\cite{schmitteckert1998} we do not observe any dependence on system
size within LDA, in agreement with the perturbative argument given 
in relation with Eq.~(\ref{eq15}).

\section{Summary}
We studied in detail the Bethe ansatz LDA for spinless fermions in one
dimension. The numerical effort of the method is comparable to the
Hartree  (or Hartree-Fock) approximation. A major improvement of LDA
with respect to the Hartree approximation is the correct prediction of
a non-charge-ordered groundstate for a large range of parameters.
Both the static and the dynamic density response functions agree
reasonably well with the exact results. In particular, for low 
density and $q < 2 k_F$, the static susceptibility obtained within LDA is almost
indistinguishable from the exact one. In the dynamic case an
impressive agreement for the velocity of collective charge
excitations is found.

On the other hand, the LDA does not capture the fact that the
system is critical with respect to a charge-ordering phase transition. Whereas
the exact susceptibility has a power law singularity at
$q= 2k_F$, the LDA susceptibility remains either finite or the system
develops charge ordering for very large system size.
As a consequence physical quantities that are related to the $2 k_F$-periodic 
charge oscillations are described incorrectly within the
local density approximation. Examples are the Friedel oscillations around
a defect, the interaction-renormalization of the reflection and
transmission probability (and therefore the conductance) and the Drude
weight.

We conclude that for applications of DFT to
one-dimensional systems improved exchange-correlation functionals
are required.

\begin{acknowledgments}
This work was supported by the Deutsche Forschungsgemeinschaft 
through SFB 484.
\end{acknowledgments}

\end{document}